\documentclass{iau}

\usepackage{amsmath}
\usepackage{graphicx}
\usepackage{multirow}
\usepackage{pifont}
\newcommand{\cmark}{\ding{51}}%
\newcommand{\xmark}{\ding{55}}%

\begin{document}

\lefttitle{Michal Zaja\v{c}ek et al.}
\righttitle{Proceedings of the International Astronomical Union}

\jnlPage{1}{7}
\jnlDoiYr{2024}
\doival{10.1017/xxxxx}

\aopheadtitle{Proceedings IAU Symposium}
\editors{eds.}

\title{Revealing EMRI/IMRI candidates with quasiperiodic ultrafast outflows}

\author{Michal Zaja\v{c}ek$^1$, Petra Suková$^2$, Vladimír Karas$^2$, Dheeraj R. Pasham$^3$, Francesco Tombesi$^{4,5,6,7,8}$, Petr Kurfürst$^1$, Henry Best$^1$, Izzy Garland$^1$, Matúš Labaj$^1$, Monika Pikhartová$^1$}
\affiliation{$^1$Department of Theoretical Physics and Astrophysics, Faculty of Science, Masaryk University, Kotlá\v{r}ská~2, 611 37 Brno, Czech Republic\\
$^2$Astronomical Institute of the Czech Academy of Sciences, Bo\v{c}ní II 1401, 141 00 Prague, Czech Republic\\$^3$Kavli Institute for Astrophysics and Space Research, Massachusetts Institute of Technology, Cambridge, MA 02139, USA\\$^4$Physics Department, Tor Vergata University of Rome, Via della Ricerca Scientifica 1, 00133 Rome, Italy\\
$^5$INAF Astronomical Observatory of Rome, Via Frascati 33, 00040 Monte Porzio Catone, Italy\\
$^6$INFN—Rome Tor Vergata, Via della Ricerca Scientifica 1, 00133 Rome, Italy\\
$^7$Department of Astronomy, University of Maryland, College Park, MD 20742, USA\\
$^8$NASA Goddard Space Flight Center, Code 662, Greenbelt, MD 20771, USA}

\begin{abstract}
The first detection of the quasiperiodic ultrafast outflow in the ASASSN-20qc system  was reported by Pasham et al. (2024). The outflow is revealed in the soft X-ray spectra as an absorption feature, which is enhanced periodically every $ \sim 8.3$ days. The repetitive nature of the ultrafast outflow is tentatively explained by an orbiting massive perturber, possibly an intermediate-mass black hole (IMBH), trajectory of which is inclined with respect to the accretion flow around the primary supermassive black hole (SMBH). In this scenario, the orbiting body pushes the disc gas into the outflow funnel, where it is accelerated by the ordered magnetic field (Sukov\'a et al. 2021). Quasiperiodic ultrafast outflows (a.k.a. QPOuts) are thus a novel phenomenon that can help reveal new extreme-/intermediate-mass ratio inspiral (EMRI/IMRI) candidates. These then would be prime candidate sources for a simultaneous detection and monitoring in electromagnetic as well as gravitational wave domains.
\end{abstract}


\maketitle

\section{Quasiperiodic ultrafast outflows (QPOuts) in galactic nuclei}

When searching for electromagnetic counterparts of large mass-ratio binary systems, in which a smaller body (a star or a compact object) orbits the supermassive black hole (hereafter SMBH) in a galactic nucleus, the focus has mainly been on quasiperiodic flares or outbursts \citep{2019Natur.573..381M,2023ApJ...957...34L,2024SSRv..220...29Z,2024ApJ...963L...1L,2024arXiv241104592S,2025arXiv250119365Z}, hence significant increases in the continuum emission of the accreting nuclear sources that are of a deterministic rather than a stochastic variability nature. One of the most monitored candidate SMBH-binary systems is the blazar OJ287, which is characterized by optical flares repeating every $\sim 11-12$ years \citep{1988ApJ...325..628S} and potentially also radio flares repeating every $\sim 20-30$ years \citep{2018MNRAS.478.3199B,2023ApJ...951..106B}. Recently, quasiperiodic X-ray eruption sources \citep[QPEs; ][]{2019Natur.573..381M,2021Natur.592..704A} raised a lot of interest since some of the interpretations of QPEs propose that they could be electromagnetic counterparts of extreme-mass/intermediate-mass ratio inspiral systems \citep[EMRIs/IMRIs;][]{2023A&A...675A.100F,2024MNRAS.532.2143K}, which are expected to be detected using low-frequency gravitational-wave observations (between $\sim 0.1$ mHz and $\sim 1$ Hz), such as Laser Interferometer Space Antenna \citep[LISA; ][]{2012CQGra..29l4016A}. However, other interpretations of QPEs have proposed that they could be caused by accretion-disc instabilities \citep{2020A&A...641A.167S,2023A&A...672A..19S} or by orbiting non-degenerate stars on mildly eccentric orbits \citep{2023MNRAS.524.6247L}, in which case the star would be tidally disrupted before reaching radii where it would be detectable using gravitational-wave observatories \citep[see, however,][for the analysis of a stripped subgiant star orbiting the SMBH and becoming a loud gravitational-wave transient in the LISA band]{2025arXiv250321995O}.

Recently, \citet{2021ApJ...917...43S} pointed out that when a secondary, smaller body with an influence radius on the order of a gravitational radius ($r_{\rm g}=GM_{\bullet}/c^2$) of the primary supermassive black hole traverses through the accretion flow at a higher inclination, not only is the accretion rate periodically perturbed but also the outflow rate. This is caused by the fact that the gas around the perturber within the influence radius is dragged along its orbit and starts comoving with the body. The influence radius could be set by the hydrodynamic ram-pressure balance (stagnation radius for wind-blowing stars or magnetized objects) or by the gravitational capture radius (see also more discussion and estimates in Section~\ref{sec_imbh}). As the perturber rises above the accretion flow, so does a fraction of the matter from the disc. Due to the ordered poloidal magnetic field, this matter is accelerated away from the SMBH and eventually moves at relativistic velocities of $\sim 0.2-0.3$ of the light speed. Moreover, the periodicity in the outflow rate was found to be significant even if the periodicity in the inflow rate was rather weak or insignificant, for instance due to a small cross-section of the perturber or its larger distance from the SMBH. If the increase in the outflow rate is associated with a partially ionized gas, this can observationally be manifested by quasiperiodic enhancements in the absorption of the underlying accretion-disc continuum emission. Hence, instead of recurrent peaks in the continuum, one would detect repetitive drops at specific energies. This is quite a different, yet realistic, concept in comparison to the canonical view of how secondary bodies surrounding SMBHs can interact with the circumnuclear gaseous-dusty medium.

The series of perturber-accretion disc interactions studied by \citet{2021ApJ...917...43S} using 2D and 3D general relativistic magnetohydrodynamic (GRMHD) simulations turned out to be useful for the interpretation of the peculiar source associated with the tidal disruption event (TDE) revealed by the optical flare ASASSN-20qc \citep{2024SciA...10J8898P}. The optical flare ASASSN-20qc/Gaia21alu/AT2020adgm detected on December 20, 2020 has been associated with the galaxy at $z=0.056$ (250 Mpc). For this galaxy, the spectroscopic measurements using broad emission lines as well as the broad-band photometry imply the SMBH mass of $M_{\bullet}=3^{+5}_{-2}\times 10^7\,M_{\odot}$, i.e. we consider the range of $M_{\bullet}=10^7-10^8\,M_{\odot}$ for further estimates. Based on eROSITA non-detections of the source in January and July 2020, the upper limit on the X-ray luminosity $L_{\rm X}\lesssim 6 \times 10^{40}\,{\rm erg\,s^{-1}}$ implies the Eddington ratio (relative accretion rate) of $\dot{m}=\dot{M}/\dot{M}_{\rm Edd}\lesssim 5\times 10^{-4}$ before the optical outburst, where we assumed the radiative efficiency of $\eta_{\rm rad}=0.1$ and the bolometric correction of $\kappa_{\rm bol}=10$. Therefore, the inner accretion flow was initially optically thin and geometrically thick hot flow dominated by advection (advection-dominated accretion flow or ADAF), which transitioned into the thin, optically thick disc at $r_{\rm ADAF}\gtrsim 2200\,\alpha_{0.1}^{0.8} \beta^{-1.08} \dot{m}_{0.0005}^{-0.53}\,r_{\rm g}$ , where the limiting radius is set by evaporation due to heat conduction between the standard cold disc and hot two-temperature corona \citep{2004A&A...428...39C}. The quantities involved in $r_{\rm ADAF}$ include the viscosity parameter $\alpha$ (scaled to 0.1) and the magnetization parameter $\beta\equiv P_{\rm g}/(P_{\rm g}+P_{\rm m})$, which stands for the ratio between the gas pressure and the total pressure ($\beta=1$ for the negligible magnetic field). The presence of the standard disc on larger scales is revealed by the detection of broad emission lines of H$\alpha$ and H$\beta$, which are dynamically related to the thin disc \citep[broad-line emitting clouds form a flattened structure; there is a significant correlation between the broad-line radius and the disc monochromatic luminosity;][]{2011A&A...525L...8C}. 

About 52 days after the optical outburst, the prominent increase in the X-ray emission was detected by the \textit{Swift telescope}, reaching the 0.3-1.1 keV luminosity of $5\times 10^{43}\,{\rm erg\,s^{-1}}$. Hence, the source soft X-ray luminosity increased by more than three orders of magnitude. At this time, the source started to be intensively monitored by the \textit{NICER} and the spectra were also obtained using the \textit{XMM-Newton telescope}. After about 100 days, the X-ray luminosity of the source dropped by two orders of magnitude to $\sim 3 \times 10^{41}\,{\rm erg\,s^{-1}}$.

\begin{figure}
    \centering
    \includegraphics[width=0.9\textwidth]{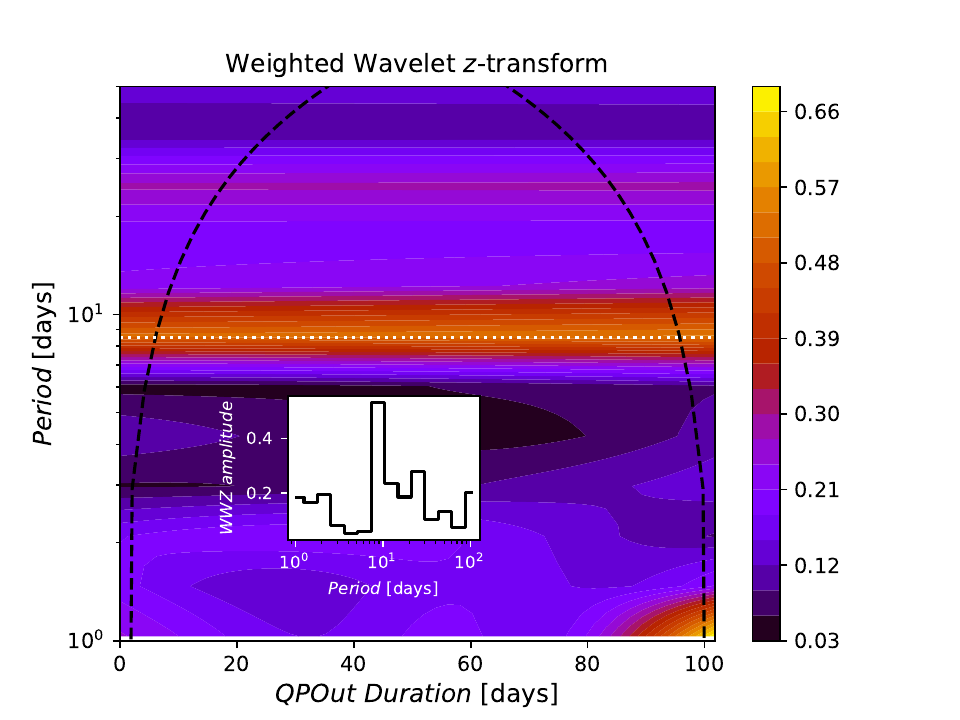}
    \caption{The periodicity determination of the quasiperiodic ultrafast outflow (QPOut) for ASASSN-20qc using the Weighted Wavelet $z$-transform (WWZ). The figure shows the WWZ amplitude as a function of both the QPOut duration and the period (both in days). The figure inset depicts the WWZ amplitude as a function of the period approximately in the middle of QPout duration. For the whole QPOut duration, there is a periodicity peak at $\sim 8.5$ days (white dotted horizontal line), which is consistent with both the Lomb-Scargle periodogram as well as the phase-minimization technique \citep{2024SciA...10J8898P}. The dashed black line represents the cone of incidence of the dataset. The colour-axis on the right represents the colour-coding for the WWZ amplitude.}
    \label{fig_wwz}
\end{figure}

During the X-ray emission peak, the X-ray spectra were characterized by a thermal accretion-disc emission between 0.30 and 0.55 keV with $kT_{\rm disc}\sim 85\,{\rm eV}$ ($T_{\rm disc}\sim 10^6\,{\rm K}$) with a broad absorption feature between 0.75 and 1.00 keV. This broad absorption feature is consistent with an ultrafast outflow moving at the velocity of $\sim 0.3c$. While the thermal X-ray continuum variability ($0.30-0.55$ keV) is clearly stochastic, the variability of the absorption band ($0.75-1.00$ keV) or rather the ratio of the flux at $0.75-1.00$ keV to the flux at $0.30-0.55$ keV, which is denoted as the \textit{outflow deficit ratio} (ODR$\equiv F_{\rm 0.75-1.00}/F_{\rm 0.30-0.55}$)\footnote{The minimum of the ODR corresponds to the enhancement of the absorbing column density, while the maximum to the decrease in the amount of the absorbing material along the line of sight.} shows a significant periodic behaviour with the period of $8.3$ days \citep{2024SciA...10J8898P}. This period is recovered when applying the Lomb-Scargle periodogram, the phase-dispersion minimization, and the weighted wavelet $z$-transform (see Fig.~\ref{fig_wwz}) to the temporal evolution of the ODR during 12 cycles, which confirms the robustness of the QuasiPeriodic ultrafast Outflow or the \textit{QPOut} feature in this source.  

\section{Dynamical interpretation}

To interpret the periodic ultrafast outflow that is responsible for the absorption at 0.75-1.00 keV, different scenarios are considered, ranging from purely geometric effects, accretion-disc instabilities, to orbiting bodies. In addition to the periodicity, one has to consider basic properties of the outflow as inferred from photoionization models of the X-ray spectra. During the ODR minima, the column density is larger ($\log{[N_{\rm h} ({\rm cm^{-2}})]}\sim 22$) in comparison with the ODR maxima ($\log{[N_{\rm h} ({\rm cm^{-2}})]}\sim 21$). The ODR minima are also characterized by a larger ionization parameter. In contrast, there is no significant difference in terms of the outflow velocity, which remains close to $\sim 0.3c$.

We provide here a brief overview including the notes and the references as well as the information whether the model is (dis)favoured considering all the observables: 

\begin{itemize}
    \item {\bf inner disc precession:} we would expect modulations of the continuum flux with the changes of the viewing angle during the precession period, i.e. $F_{\rm X} \propto \cos{\iota}$, see also \citet{2024Natur.630..325P}; \underline{disfavoured} \textcolor{red}{\xmark},
    \item {\bf clumpy disc wind:} we would expect stochastic absorption variability, while significant periodicity of $8.3$ days was detected; \underline{disfavoured} \textcolor{red}{\xmark},
    \item {\bf X-ray reflection contribution:} it can be ruled out based on the absence of the hard X-ray contribution (hot corona); \underline{disfavoured} \textcolor{red}{\xmark},
    \item {\bf magnetically arrested disc (MAD) with outflowing plasmoids due to plasmoid-mediated reconnection events:} see e.g. \citet{2022ApJ...924L..32R}; however, the outflow rate is rather stochastic and the outflow velocities are expected to exhibit a higher velocity component in comparison with the perturber-induced outflow ($\gtrsim 0.5c$), see also \citet{2023arXiv231204149S} for the comparison; \underline{disfavoured} \textcolor{red}{\xmark},   
    \item {\bf quasi-periodic X-ray eruptions (QPEs):} although the mechanism behind the quasiperiodic nuclear phenomenon of QPEs \citep{2019Natur.573..381M,2021Natur.592..704A} is still under debate, it is characterized by rapid, short increases in the soft X-ray thermal continuum emission by $1-2$ orders of magnitude. This is clearly not the case for ASASS-20qc since the thermal continuum variability is stochastic without large-amplitude flares; \underline{disfavoured} \textcolor{red}{\xmark},    
    \item {\bf repeating partial tidal disruption events (rpTDEs):} rpTDEs are characterized by periodic optical/UV continuum flares recurring with a longer period of $\sim 100$ days, see e.g. \citet{2022ApJ...926..142P}, which is not the case for ASASSN-20qc; \underline{disfavoured} \textcolor{red}{\xmark}, 
    \item {\bf outflows driven by radiation-pressure instability:} there should be flares in the continuum with a certain duty cycle that depends on a number of parameters. Such flares in the continuum are not observed for ASASS-20qc. In addition, the relative accretion rate is lower, even during the epoch with the increased X-ray emission ($\dot{m}\lesssim 0.04$), than the value expected for the radiation-pressure instability to operate \citep[$\dot{m}_{\rm RPI} \gtrsim 0.14(\alpha/0.1)^{41/29} (M_{\bullet}/10^7\,M_{\odot})^{-1/29}$][]{2020A&A...641A.167S}; \underline{disfavoured} \textcolor{red}{\xmark},   
    \item {\bf periodic ultrafast outflow launched by an orbiting perturber:} a massive perturber, which is likely an intermediate-mass black hole, can periodically enhance the outflow rate as it orbits around the SMBH. Furthermore, the launched gas clumps are futher accelerated to velocities that are comparable to the ones inferred for ASASSN-20qc \citep[see also][]{2021ApJ...917...43S}, without a high-velocity part present for the outflow-rate peaks in the MAD model. This model can, in principle, capture all the main characteristics of the QPOuts in the ASASSN-20qc host, as we address more in the following section;  \underline{favoured} \textcolor{green}{\cmark}.
\end{itemize}    

\section{QPOuts induced by an orbiting body: black holes vs. stars}
\label{sec_imbh}

\begin{figure}
    \centering
    \includegraphics[width=0.9\textwidth]{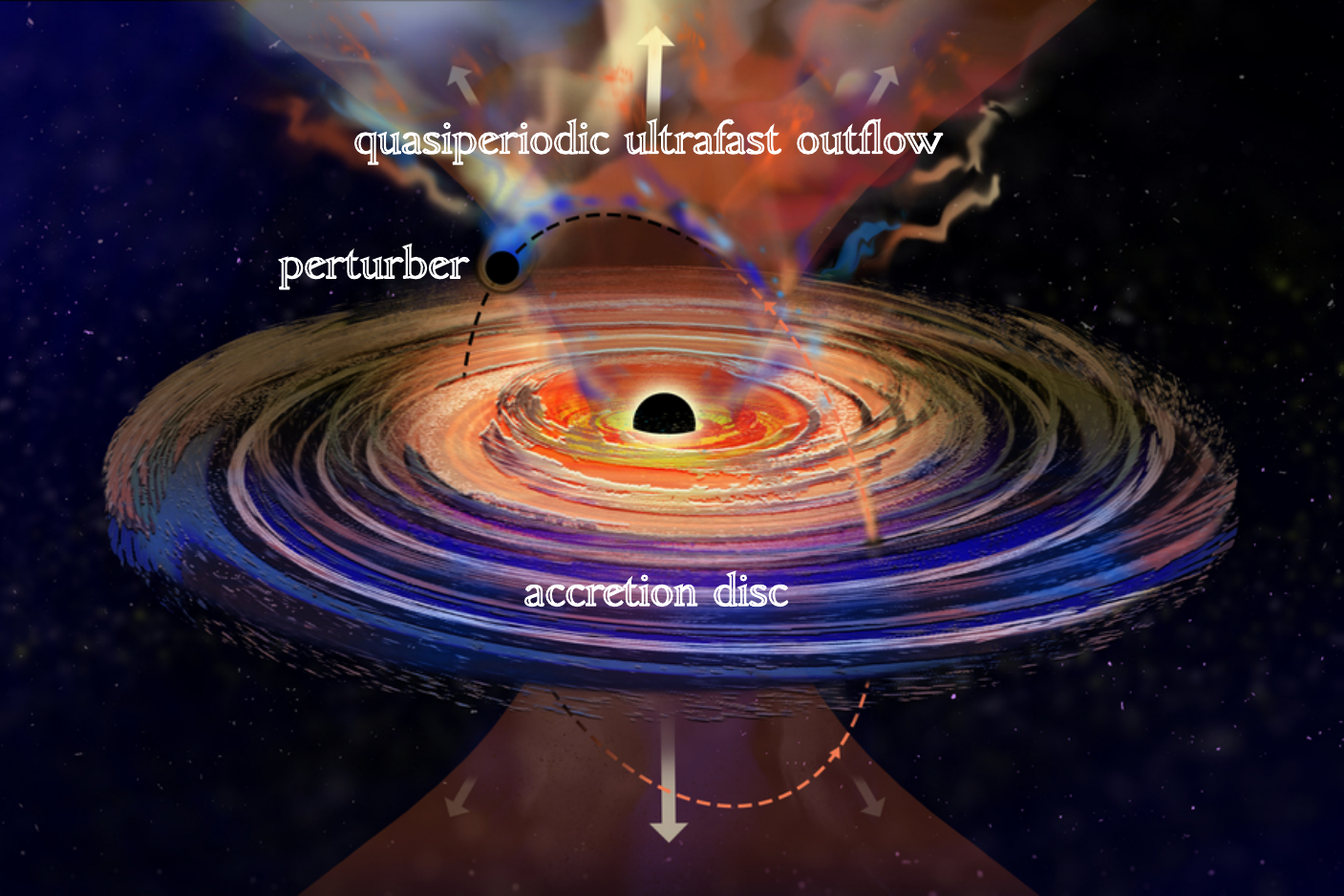}
    \caption{Illustration of the generation of the quasiperiodic ultrafast outflow (QPOut) as the perturbing body orbits the SMBH. During each orbit, the perturber crosses the disc twice but only when it goes above the disc plane towards the observer, the pushed-out gas can block the underlying continuum emission, which results in the detected absorption feature in the soft X-ray spectrum. Image credit: Jose-Luis Olivares, MIT.}
    \label{fig_QPOut_illustration}
\end{figure}

In the model setup, we assume that the clumps that are further accelerated by the ordered, poloidal magnetic field can cause significant absorption of the underlying thermal continuum only when the orbiting body rises above the thermally emitting disc. Along with the perturber, the gas is pushed towards the observer and blocks a fraction of the thermal continuum, which presumably originates in the compact disc formed due to the TDE (see Fig.~\ref{fig_QPOut_illustration} for an artistic illustration of the setup). Although the gas flow circularization of the stellar material following the TDE is rather complex, we adopt the scenario where the X-ray emitting accretion disc forms on the length-scale given by the tidal disruption radius of the star \citep{2023ApJ...957...34L}. The disc formation revealed by the increase in the X-ray emission is expected to be delayed with respect to the initial optical flare by the timescale characteristic of the flow circularization \citep[free-fall timescale, viscous timescale;][]{2017ApJ...837L..30P,2024SSRv..220...29Z}.

\begin{figure}
    \centering
    \includegraphics[width=0.9\textwidth]{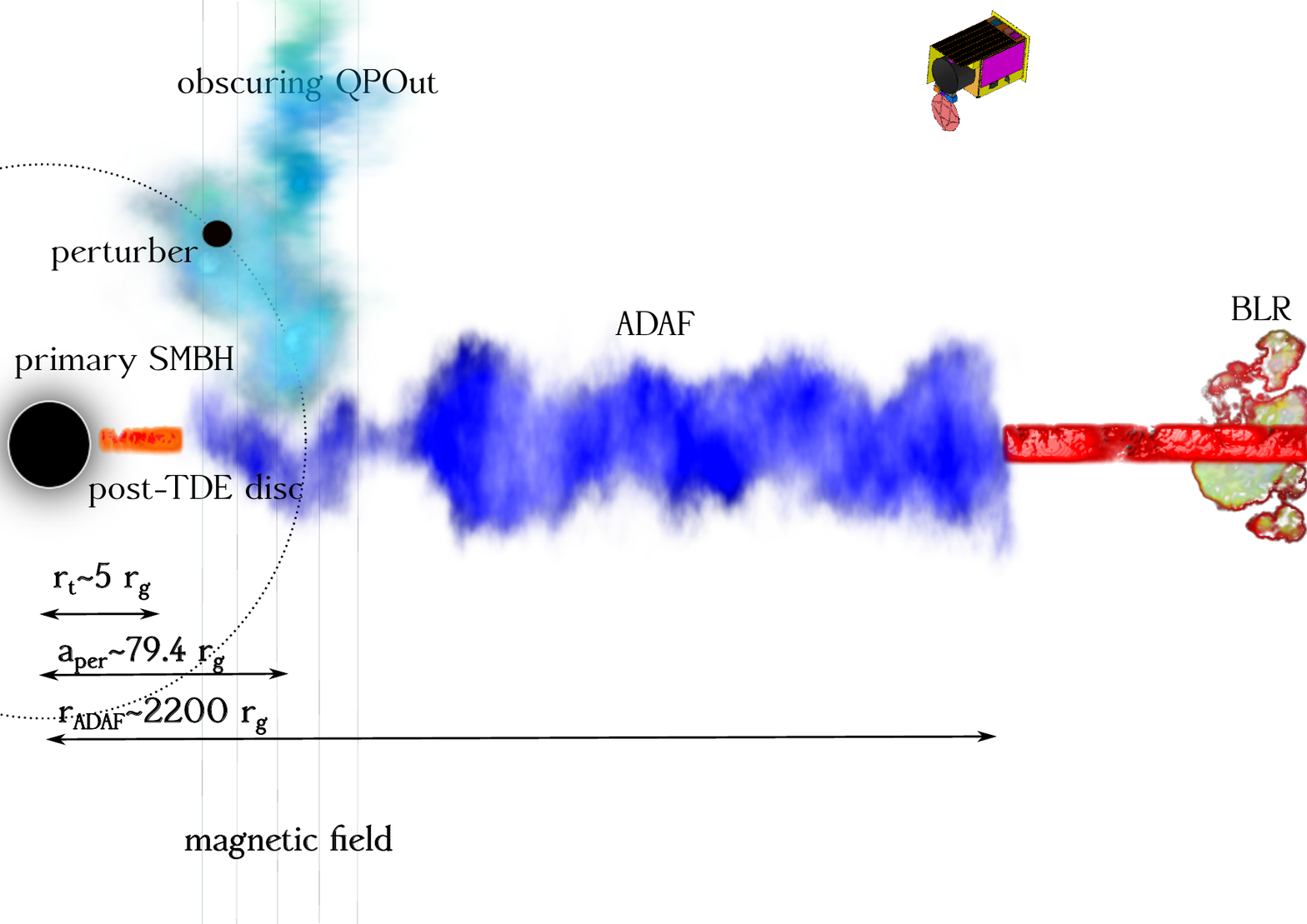}
    \caption{Illustration of the complex accretion flow in the ASASSN-20qc host. Due to the previous low-luminosity phase, the advection-dominated accretion flow (ADAF) is expected to extend over a large range of radii. At the outer parts of the accretion flow (at $r_{\rm ADAF}$), the disc condenses into the standard disc, which is hinted by the presence of broad lines (BLR). After the passing-by star is tidally disrupted, a colder and compact accretion disc forms within $r_{\rm t}$ that emits thermal X-rays. A perturber orbiting at $a_{\rm per}$ crosses an accretion flow, presumably where the ADAF still exists, and ejects blobs of gas that obscure the thermal X-ray emission. }
    \label{fig_composite_AF}
\end{figure}

In this way, the detected QPOut period approximately coincides with the orbital period of the body, $P_{\rm QPOut}\simeq P_{\rm orb}\sim 8.0$ days, where we took into account the redshift of the source to get the rest-frame period of the QPOuts/perturber. Scaling the primary SMBH mass to $10^{7.5}\,M_{\odot}$, we can constrain the expected distance/semi-major axis of the perturbing body $a_{\rm per}$ in gravitational radii of the primary SMBH,
\begin{equation}
    \frac{a_{\rm per}}{r_{\rm g}}\simeq 79.4 \left(\frac{P_{\rm orb}}{8\,{\rm days}} \right)^{2/3} \left(\frac{M_{\bullet}}{10^{7.5}\,M_{\odot}} \right)^{-2/3}\,.
    \label{eq_distance_per}
\end{equation}
Since before the TDE-like optical outburst the ASASS-20qc host was low-luminous, the inner accretion flow was most likely a hot ADAF up to the distance of a few 1000 $r_{\rm g}$. The TDE brought stellar material, from which a compact, thermally emitting disc formed on the scale of a tidal radius \citep{2023ApJ...957...34L},
\begin{equation}
\frac{r_{\rm t}}{r_{\rm g}}\simeq 5\left(\frac{R_{\star}}{1\,R_{\odot}} \right) \left( \frac{M_{\bullet}}{10^{7.5}\,M_{\odot}}\right)^{-2/3} \left( \frac{m_{\star}}{1\,M_{\odot}}\right)^{-1/3}\,,
\end{equation}
which is about one order of magnitude less than the orbital distance of the perturber, assuming the stellar mass of $m_{\star}=1\,M_{\odot}$. Hence, it is quite plausible that the orbiting body crosses a hot flow rather than the thin disc, and the ejected gas blobs periodically obscure the thermal continuum emitted by the inner post-TDE disc as is shown in Fig.~\ref{fig_composite_AF}, which illustrates the likely distribution of the circumnuclear material within the composite accretion flow in the ASASS-20qc host.

Different options have been discussed in the literature regarding the nature of the orbiter: a giant star with an extended atmosphere that becomes gradually ablated by the surrounding environment \citep{1992AIPC..254..564Z,2004MNRAS.354.1177S,2023ApJ...957...34L}, confined core-less clouds \citep{1993ApJ...412L..17M}, a magnetic star (e.g.\ a strongly magnetized neutron star or Ap star), where the interaction radius is defined by the extent of its magnetosphere that can reach well above the direct geometrical cross-section \citep{1992ans..book.....L,2012MNRAS.420..810T,2015AcPol..55..203Z,2017CoSka..47..124K,2018PhRvD..98h4055K}, and an intermediate-mass black hole (IMBH) that is the most compact possibility, whereas the interaction cross-section is sufficiently large due to its large mass (i.e.\ super-solar, albeit less than SMBH; \citeauthor{2024SciA...10J8898P} \citeyear{2024SciA...10J8898P}).

The effective cross-sectional area for the interaction between the body of a tentative perturber and the gaseous/dusty medium is the crucial aspect that defines the efficiency of the mutual interaction between the accretion flow and the transiting body. In fact, the effects of magnetic fields with an organized (large-scale) component permeating the accretion disc corona or a jet play an important role (as suggested in various simulations and indicated by polarimetric observations). Until the imaging resolution reaches $\sim$ microarcsecond level \citep[as in e.g.][]{2018A&A...618L..10G,2021ApJ...910L..13E,2024ApJ...964L..25E}, the most likely scenario has to be inferred from supplementary evidence consistent with the timing and spectral properties of the system under discussion.

In the context of ASASSN-20qc we have pursued the latter cited scheme (IMBH perturber) as the most appealing possibility that is in agreement with currently available observational evidence. To infer the required influence radius of the perturber, we consider the column density of the outflow $N_{\rm h}\sim 2 \times 10^{22}\,{\rm cm^{-2}}$ during the ODR minima. Here we consider the interaction of the orbiter with the ADAF flow with the number density $n_{\rm ADAF}$. The length-scale of the ejected clumps can be estimated as $R_{\rm clump}\approx f_{\rm g} N_{\rm h}/n_{\rm ADAF}$, where $f_{\rm g}$ is the geometrical/dynamical factor related to the clump shape and evolution shortly after the ejection. Considering the number density for the ADAF at a given distance, $n_{\rm ADAF}\simeq 5.02 \times 10^9\,(\alpha/0.1)^{-1}(M_{\bullet}/10^{7.5}\,M_{\odot})^{-1}(\dot{m}/0.01)(r/79.4\,r_{\rm g})^{-1}{\rm cm^{-3}}$ \citep{2014ARA&A..52..529Y}, we can estimate the typical clump radius in gravitational radii,
\begin{equation}
    R_{\rm clump}\approx 0.9 \left(\frac{N_{\rm h}}{2 \times 10^{22}\,{\rm cm^{-2}}} \right) \left(\frac{n_{\rm ADAF}}{5 \times 10^{9}\,{\rm cm^{-3}}} \right)^{-1} \left(\frac{M_{\bullet}}{10^{7.5}\,M_{\odot}} \right)^{-1} r_{\rm g}\,.
    \label{eq_clump_radius}
\end{equation}
In order to eject clumps of the size $R_{\rm clump}$, the influence radius of the perturber should be comparable. Therefore, for the following estimates, we set $R_{\rm inf}\approx 1\,r_{\rm g}$. The length-scale of $R_{\rm inf}$ close to the gravitational radius is independently motivated by the results of GRMHD simulations that were tailored to reproduce the outflow/inflow rate ratio in ASASSN-20qc \citep{2024SciA...10J8898P}. Based on the analysis of GRMHD simulation time series of the inflow/outflow rates, a comparable length-scale of the influence radius of the perturber was inferred by \citet{2021ApJ...917...43S}.

The mass of the compact (inert) orbiter can be constrained by the tidal (Hill) radius, $R_{\rm Hill}=r_{\rm per}[m_{\rm per}/(3M_{\bullet})]^{1/3}$ which stands for the distance range from the body where it dominates over the SMBH gravitational influence. This yields the Hill mass of,
\begin{align}
    m_{\rm per}^{\rm Hill} &=\frac{12 \pi^2 G^2}{c^6} \left(\frac{R_{\rm inf}}{r_{\rm g}} \right)^3 \frac{M_{\bullet}^3}{P_{\rm orb}^2}\,\notag\\
    &\simeq 189\,\left(\frac{R_{\rm inf}}{1\,r_{\rm g}} \right)^3 \left(\frac{M_{\bullet}}{10^{7.5}\,M_{\odot}} \right)^3 \left(\frac{P_{\rm orb}}{8\,\text{days}} \right)^{-2} M_{\odot}\,.
    \label{eq_HilL_mass}
\end{align}
Another constraint can be obtained from the Bondi (gravitational capture) radius, $R_{\rm B}=2Gm_{\rm per}/(v_{\rm rel}^2+c_{\rm s}^2)$, which expresses the distance from the perturber, where it dominates over the thermal motion of the surrounding ADAF,
\begin{align}
    m_{\rm per}^{\rm Bondi}&=\frac{(2c^2+f_{\rm s})G^{2/3}(4\pi^2)^{1/3}}{2c^4}\left(\frac{R_{\rm inf}}{r_{\rm g}} \right)\frac{M_{\bullet}^{5/3}}{P_{\rm orb}^{2/3}}\,\notag\\
    &\simeq 4.6 \times 10^5\,\left(\frac{R_{\rm inf}}{1\,r_{\rm g}} \right) \left(\frac{M_{\bullet}}{10^{7.5}\,M_{\odot}} \right)^{5/3} \left(\frac{P_{\rm orb}}{8\,\text{days}} \right)^{-2/3} M_{\odot}\,, 
\end{align}
where $f_{\rm s}\approx 2.8 \times 10^{20}\,{\rm cm^2\,s^{-2}}$ is the normalization constant related to the sound speed in the ADAF \citep{2014ARA&A..52..529Y}. For the more direct association with the GRMHD simulations of the accretion-disc perturbation \citep{2021ApJ...917...43S}, the influence radius is rather associated with the comoving radius, i.e. where the surrouding gas is gravitationally captured, dragged, and starts comoving with the perturber. To the first approximation, this radius is comparable to the Bondi-Hoyle-Lyttleton radius modulated by the factor of the order of unity that depends on the Mach number of the perturber \citep{1999ApJ...513..252O,2021ApJ...917...43S}. Hence, the length-scales associated with the gravitational attraction of the orbiting body imply that it is a black hole in the intermediate-mass range between $\sim 10^2\,M_{\odot}$ and $\sim 10^5\,M_{\odot}$.

An independent constraint or rather an upper limit on the IMBH mass is given by the gravitational-merger timescale $\tau_{\rm gw}$ \citep{1964PhRv..136.1224P}. It can be argued that $\tau_{\rm gw}$ should be at least comparable to or longer than the timescale related to the average TDE rate per galaxy, which is $\tau_{\rm TDE}\gtrsim 10^4$ years \citep{2016MNRAS.455..859S}. Otherwise the occurrence of the TDE during the lifetime of the SMBH--IMBH system would be very unlikely. This statistical argument yields the upper limit on the IMBH mass,
\begin{align}
    m_{\rm per}&\lesssim \frac{5c^5}{256 (4\pi^2)^{4/3}G^{5/3}}  \frac{P_{\rm orb}^{8/3}}{M_{\bullet}^{2/3}\tau_{\rm TDE}}\,\notag\\
    &\simeq 12100\,\left(\frac{P_{\rm orb}}{8\,\text{days}} \right)^{8/3} \left(\frac{M_{\bullet}}{10^{7.5}\,M_{\odot}} \right)^{-2/3} \left(\frac{\tau_{\rm TDE}}{10^4\,\text{years}} \right)^{-1}M_{\odot}\,.
    \label{eq_merger_mass}
\end{align}

The constraints on $\tau_{\rm TDE}$ for the ASASSN-20qc host galaxy, which can be inferred from the photomery and the surface-brightness distribution of the stellar population \citep[recent starburst; a core-like or a cusp-like nuclear stellar cluster;][]{2016ApJ...825L..14S}, could thus help constrain the IMBH mass range. In particular, the indication for a longer $\tau_{\rm TDE}$ would result in narrowing down the mass range for the IMBH.

The constraints given by the required influence radius in combination with the gravitational effect on the surrounding gas, Eq.~\eqref{eq_HilL_mass}, and the long enough gravitational merger timescale of the SMBH-IMBH system, Eq.~\eqref{eq_merger_mass}, yield the IMBH mass in the range between $\sim 10^2\,M_{\odot}$ and $\sim 10^4\,M_{\odot}$. Concerning an alternative explanation involving a normal star, it would need to reach a stellar radius of $R_{\star}\approx R_{\rm inf}\simeq  1\,r_{\rm g}\sim 67\,(M_{\bullet}/10^{7.5}\,M_{\odot})R_{\odot}$, hence it would need to be an evolved, late-type star. However, given the tidal (Hill) radius for $1\,M_{\odot}$ star orbiting the SMBH of $M_{\bullet}=10^{7.5}\,M_{\odot}$ every 8 days,
\begin{align}
    \frac{R_{\rm Hill}}{R_{\odot}}&\simeq \left(\frac{G}{12 \pi^2} \right)^{1/3} \frac{P_{\rm orb}^{2/3}m_{\star}^{1/3}}{R_{\odot}}\,\notag\\
    &\approx 12\left(\frac{P_{\rm orb}}{8\,\text{days}} \right)^{2/3}\left(\frac{m_{\star}}{1\,M_{\odot}}\,, \right)^{1/3}\,
\end{align}
such a large star would get disrupted during one orbital period. However, for lighter SMBHs in other galaxies, the star could be tidally stable and at the same it would have the influence radius of $\sim 1\,r_{\rm g}$. The condition $R_{\rm Hill}\gtrsim R_{\star}\sim 1\,r_{\rm g}$ yields the limiting SMBH mass $M_{\bullet}^{\rm lim}$, below which stellar and black-hole perturber regimes are both plausible, while above it, black-hole orbiters perturbing the accretion flow in a significant manner are more likely. Considering the comparable periodicity as in ASASSN-20qc, we can estimate $M_{\bullet}^{\rm lim}$ as
\begin{align}
    M_{\bullet}^{\rm lim}&\lesssim \frac{c^2}{(12 \pi^2)^{1/3} G^{2/3}} P_{\rm orb}^{2/3} m_{\star}^{1/3}\,\notag\\
    &\sim 10^{6.74}\, \left(\frac{P_{\rm orb}}{8\,\text{days}} \right)^{2/3} \left(\frac{m_{\star}}{1\,M_{\odot}} \right)^{1/3}\,M_{\odot}\,.
\end{align}

\begin{figure}
    \centering
    \includegraphics[width=\textwidth]{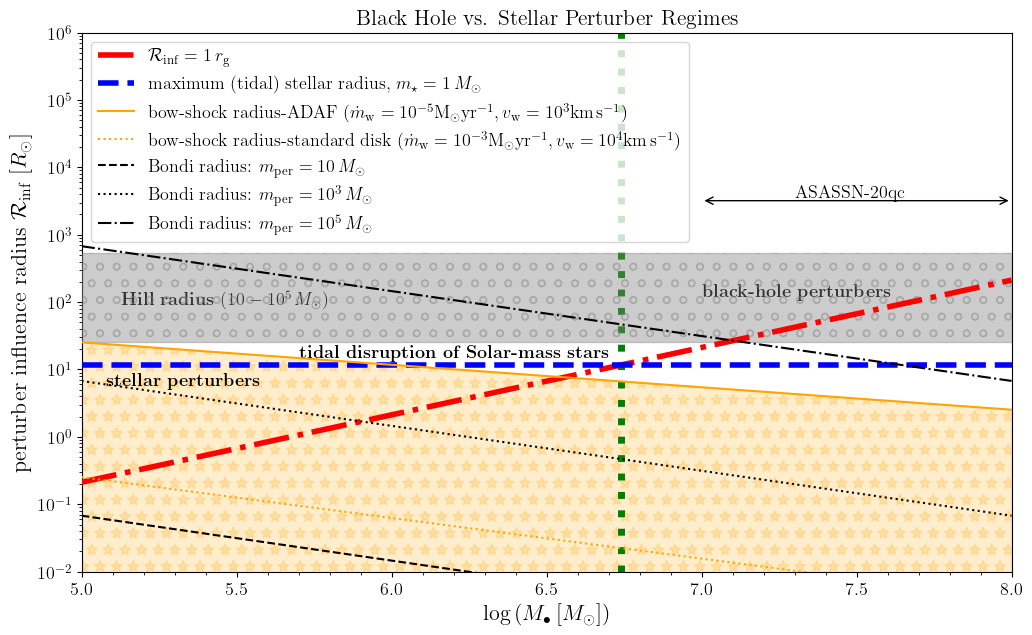}
    \caption{Influence radius of a body (a star or a black hole) expressed in Solar radii  as a function of the primary SMBH mass. The red dot-dashed line stands for $R_{\rm inf}=1\,r_{\rm g}$. The gray shaded region depicts tidal (Hill) radii of bodies in the range of $10$-$10^5\,M_{\odot}$. Black lines stand for Bondi radii of massive bodies (black holes) according to the legend. The blue dashed line represents the tidal radius of a Solar-mass star. Yellow-shaded region (and corresponding lines) pertain to bow-shock sizes produced by wind-blowing stars. The vertical dotted green line represents the SMBH mass of $M_{\bullet}=10^{6.74}\,M_{\odot}$, which separates the purely black-hole perturber regime for more massive SMBHs from the mixed stellar/black-hole perturber regime for less massive SMBHs. The orbital period of orbiting bodies is set to $P_{\rm orb}=8$ days as for ASASSN-20qc.  }
    \label{fig_star_BH_regimes}
\end{figure}

For $M_{\bullet}^{\rm lim}$, the star with the influence radius of $1\,r_{\rm g}$ would have the physical radius of $R_{\star}\sim 11.6\,R_{\odot}$. In these estimates, we have not considered hydrodynamical effects, such as strong stellar winds, however, for the accretion rate of $\dot{m}\sim0.05$, they do not yield large enough kinetic pressure to increase the influence radius of stars above their physical radii. In Fig.~\ref{fig_star_BH_regimes}, we plot the influence radius in Solar radii as a function of the SMBH mass, considering $R_{\rm inf}=1\,r_{\rm g}$. The separation of stellar and black-hole perturber regimes around $10^{6.74}\,M_{\odot}$ is depicted by plotting tidal (Hill) radius for $m_{\star}=1\,M_{\odot}$ and at the same time tidal (Hill)/Bondi radii for black-hole perturbers with $m_{\rm per}=10-10^5\,M_{\odot}$ (gray-shaded region and black lines); see also \citet{2024SciA...10J8898P} for details. For completeness, we also show expected stellar bow-shock sizes in both the ADAF and the standard disc for the parameters listed in the legend.

\section{Implications for EMRI/IMRI systems}

So far the studies of electromagnetic counterparts of EMRI/IMRI systems focused mostly on the flux density enhancements (flares, eruptions) due to inclined smaller bodies crossing an accretion disc, producing shocks emitting mostly in soft X-ray/UV domains. The seminal study of \citet{2021ApJ...917...43S} showed that perturbing bodies with a sufficient cross-section (compact remnants or stars) could actually affect the nuclear outflow rate more visibly than the inflow rate, especially if there are located at a relatively larger distance from the primary massive black hole. Observationally, such an effect can be revealed by the periodically launched ultrafast outflow (QPOut) resulting in the enhanced absorption of the underlying disc continuum, as was shown to be the likely cause of the observed spectral changes in the soft X-ray domain in the ASASSN-20qc host \citep{2024SciA...10J8898P}.

The tidal (Hill) radius of $R_{\rm inf}\approx 1\,r_{\rm g}$ puts a lower limit on the perturbing body mass of $m_{\rm per}\gtrsim 100\,M_{\odot}$. The statistical argument that the merger timescale of the system should be longer than the timescale associated with the typical TDE rate per galaxy put an upper limit on the perturber mass of $m_{\rm per}\lesssim 10^4\,M_{\odot}$. Hence, the best dynamical explanation of the QPOut in the ASASSN-20qc host appears to be an IMBH orbiting close to $\sim 100\,r_{\rm g}$ on a moderately eccentric orbit. This can address (i) the QPOut as well as (ii) no significant periodic variability of the inflow/accretion rate. 

The constraints on the perturber mass depend on the primary SMBH mass. It can be shown that for the required influence radius of $\sim 1\,r_{\rm g}$, perturbers can also be normal stars if the primary SMBH mass is $\lesssim 10^{6.74}\,M_{\odot}$. This has consequences for the interpretation of nuclear quasiperiodic photometric and spectral changes in other hosts. 

Another relevant aspect of the study of the ASASSN-20qc event are the implications for the origin of IMBHs. Since they are only very few confirmed cases of IMBHs \citep{2020ARA&A..58..257G}, the QPOuts provide a novel observational way to look for their fingerprints. The broad mass range for the putative IMBH in the ASASSN-20qc host ($10^2\,M_{\odot}\lesssim m_{\rm per} \lesssim 10^4\,M_{\odot}$) is consistent with the following three proposed scenarios for the origin of SMBH-IMBH binaries:
\begin{itemize}
    \item infall of a globular/star-forming cluster hosting an IMBH with a certain occupancy franction \citep{2022ApJ...939...97F};
    \item repetitive stellar black hole -- main-sequence star collisions in nuclear star clusters \citep{2022ApJ...929L..22R}, which can result in the IMBH with an upper limit of $10^4\,M_{\odot}$;
    \item repetitive stellar black hole mergers in nuclear star clusters \citep{2022ApJ...927..231F}. The merger products possess recoil velocities, but they can mostly be retained within massive nuclear clusters. The resulting IMBH mass is in the range of $\sim 10^3-10^4\,M_{\odot}$.
\end{itemize}

The merger rates of stellar black holes (involving both compact remnants and stars) are expected to be greatly enhanced in the zones called \textit{migration traps} where the smaller objects tend to accumulate due to the boundary between the inward and the outward migration of objects due to gas torques within the accretion disc \citep{2012MNRAS.425..460M,2014MNRAS.441..900M}. The inferred distance of the IMBH in ASASSN-20qc ($\sim 100\,r_{\rm g}$) roughly coincides with the distance range for such a migration trap created by a Type I migration mechanism \citep[$\sim 40-600\,r_{\rm g}$; ][]{2016ApJ...819L..17B} or below the inner radius of the migration trap created by the thermal torques \citep[$\sim 10^3-10^5\,r_{\rm g}$;][]{2024MNRAS.530.2114G}.

The interaction of an orbiting body with the standard accretion disc will make it aligned with the disc plane due to efficient hydrodynamical drag on the timescale of $\sim 10^4$ years \citep{1991MNRAS.250..505S,2024SciA...10J8898P}, assuming the parameters comparable to the ASASSN-20qc system. If the perturber embedded within the disc is massive enough, of the order of $10^{-2}$ of the primary SMBH mass \citep{2012ApJ...761...90G,2023MNRAS.522.2869S} and potentially even lighter \citep{2025ApJ...982L..13G}, it opens a gap in the disc. The gap presence is imprinted in the spectral energy distribution (SED) of the nuclear source, specifically it leads to flux density depressions at specific wavelengths (from the UV to the optical bands) corresponding to the distance of the perturber from the SMBH \citep{2023MNRAS.522.2869S}. In comparison with inclined orbiting bodies, the aligned, embedded perturber thus creates a quasistationary effect, which can be revealed in one single-epoch SED, i.e. with the flux density measurements obtained in a narrow time range. The gap opening can also prevent the formation of migration traps in the accretion discs and can thus suppress the hierarchical growth of black holes \citep{2025ApJ...982L..13G}. This would conversely have consequences for the SMBH-IMBH merger rates.

\section*{Acknowledgements}

We are grateful to the anonymous referee for providing constructive comments that helped to improve the manuscript. MZ, PS, VK, HB, IG, and MP acknowledge the support from the Czech Science Foundation (GA\v{C}R) Junior Star grant no. GM24-10599M.  PK and ML acknowledge the continuing support of the OPUS-LAP/GAČR-LA
bilateral research grant (2021/43/I/ST9/01352/OPUS22 and
GF23-04053L).


\begin{thebibliography}{54}
\expandafter\ifx\csname natexlab\endcsname\relax\def\natexlab#1{#1}\fi

\bibitem[{{Amaro-Seoane} {et~al.}(2012){Amaro-Seoane}, {Aoudia}, {Babak}, {Bin{\'e}truy}, {Berti}, {Boh{\'e}}, {Caprini}, {Colpi}, {Cornish}, {Danzmann}, {Dufaux}, {Gair}, {Jennrich}, {Jetzer}, {Klein}, {Lang}, {Lobo}, {Littenberg}, {McWilliams}, {Nelemans}, {Petiteau}, {Porter}, {Schutz}, {Sesana}, {Stebbins}, {Sumner}, {Vallisneri}, {Vitale}, {Volonteri}, \& {Ward}}]{2012CQGra..29l4016A}
{Amaro-Seoane}, P., {Aoudia}, S., {Babak}, S., {et~al.} 2012, Classical and Quantum Gravity, 29, 124016

\bibitem[{{Arcodia} {et~al.}(2021){Arcodia}, {Merloni}, {Nandra}, {Buchner}, {Salvato}, {Pasham}, {Remillard}, {Comparat}, {Lamer}, {Ponti}, {Malyali}, {Wolf}, {Arzoumanian}, {Bogensberger}, {Buckley}, {Gendreau}, {Gromadzki}, {Kara}, {Krumpe}, {Markwardt}, {Ramos-Ceja}, {Rau}, {Schramm}, \& {Schwope}}]{2021Natur.592..704A}
{Arcodia}, R., {Merloni}, A., {Nandra}, K., {et~al.} 2021, Nature, 592, 704

\bibitem[{{Bellovary} {et~al.}(2016){Bellovary}, {Mac Low}, {McKernan}, \& {Ford}}]{2016ApJ...819L..17B}
{Bellovary}, J.~M., {Mac Low}, M.-M., {McKernan}, B., \& {Ford}, K.~E.~S. 2016, ApJL, 819, L17

\bibitem[{{Britzen} {et~al.}(2018){Britzen}, {Fendt}, {Witzel}, {Qian}, {Pashchenko}, {Kurtanidze}, {Zaja{\v{c}}ek}, {Martinez}, {Karas}, {Aller}, {Aller}, {Eckart}, {Nilsson}, {Ar{\'e}valo}, {Cuadra}, {Subroweit}, \& {Witzel}}]{2018MNRAS.478.3199B}
{Britzen}, S., {Fendt}, C., {Witzel}, G., {et~al.} 2018, MNRAS, 478, 3199

\bibitem[{{Britzen} {et~al.}(2023){Britzen}, {Zaja{\v{c}}ek}, {Gopal-Krishna}, {Fendt}, {Kun}, {Jaron}, {Sillanp{\"a}{\"a}}, \& {Eckart}}]{2023ApJ...951..106B}
{Britzen}, S., {Zaja{\v{c}}ek}, M., {Gopal-Krishna}, {et~al.} 2023, ApJ, 951, 106

\bibitem[{{Czerny} \& {Hryniewicz}(2011)}]{2011A&A...525L...8C}
{Czerny}, B. \& {Hryniewicz}, K. 2011, A\&A, 525, L8

\bibitem[{{Czerny} {et~al.}(2004){Czerny}, {R{\'o}za{\'n}ska}, \& {Kuraszkiewicz}}]{2004A&A...428...39C}
{Czerny}, B., {R{\'o}za{\'n}ska}, A., \& {Kuraszkiewicz}, J. 2004, A\&A, 428, 39

\bibitem[{{Event Horizon Telescope Collaboration} {et~al.}(2024){Event Horizon Telescope Collaboration}, {Akiyama}, {Alberdi}, {Alef}, {Algaba}, {Anantua}, {Asada}, {Azulay}, {Bach}, {Baczko}, {Ball}, {Balokovic}, {Bandyopadhyay}, {Barrett}, {Baub{\"o}ck}, {Benson}, {Bintley}, {Blackburn}, {Blundell}, {Bouman}, {Bower}, {Boyce}, {Bremer}, {Brinkerink}, {Brissenden}, {Britzen}, {Broderick}, {Broguiere}, {Bronzwaer}, {Bustamante}, {Byun}, {Carlstrom}, {Ceccobello}, {Chael}, {Chan}, {Chang}, {Chatterjee}, {Chatterjee}, {Chen}, {Chen}, {Cheng}, {Cho}, {Christian}, {Conroy}, {Conway}, {Cordes}, {Crawford}, {Crew}, {Cruz-Osorio}, {Cui}, {Dahale}, {Davelaar}, {De Laurentis}, {Deane}, {Dempsey}, {Desvignes}, {Dexter}, {Dhruv}, {Dihingia}, {Doeleman}, {Dougal}, {Dzib}, {Eatough}, {Emami}, {Falcke}, {Farah}, {Fish}, {Fomalont}, {Ford}, {Foschi}, {Fraga-Encinas}, {Freeman}, {Friberg}, {Fromm}, {Fuentes}, {Galison}, {Gammie}, {Garc{\'\i}a}, {Gentaz}, {Georgiev}, {Goddi}, {Gold}, {G{\'o}mez-Ruiz}, {G{\'o}mez}, {Gu},
  {Gurwell}, {Hada}, {Haggard}, {Haworth}, {Hecht}, {Hesper}, {Heumann}, {Ho}, {Ho}, {Honma}, {Huang}, {Huang}, {Hughes}, {Ikeda}, {Impellizzeri}, {Inoue}, {Issaoun}, {James}, {Jannuzi}, {Janssen}, {Jeter}, {Jiang}, {Jim{\'e}nez-Rosales}, {Johnson}, {Jorstad}, {Joshi}, {Jung}, {Karami}, {Karuppusamy}, {Kawashima}, {Keating}, {Kettenis}, {Kim}, {Kim}, {Kim}, {Kim}, {Kino}, {Koay}, {Kocherlakota}, {Kofuji}, {Koch}, {Koyama}, {Kramer}, {Kramer}, {Kramer}, {Krichbaum}, {Kuo}, {La Bella}, {Lauer}, {Lee}, {Lee}, {Leung}, {Levis}, {Li}, {Lico}, {Lindahl}, {Lindqvist}, {Lisakov}, {Liu}, {Liu}, {Liuzzo}, {Lo}, {Lobanov}, {Loinard}, {Lonsdale}, {Lowitz}, {Lu}, {MacDonald}, {Mao}, {Marchili}, {Markoff}, {Marrone}, {Marscher}, {Mart{\'\i}-Vidal}, {Matsushita}, {Matthews}, {Medeiros}, {Menten}, {Michalik}, {Mizuno}, {Mizuno}, {Moran}, {Moriyama}, {Moscibrodzka}, {Mulaudzi}, {M{\"u}ller}, {M{\"u}ller}, {Mus}, {Musoke}, {Myserlis}, {Nadolski}, {Nagai}, {Nagar}, {Nakamura}, {Narayanan}, {Natarajan}, {Nathanail}, {Fuentes},
  {Neilsen}, {Neri}, {Ni}, {Noutsos}, {Nowak}, {Oh}, {Okino}, {Olivares}, {Ortiz-Le{\'o}n}, {Oyama}, {{\"O}zel}, {Palumbo}, {Paraschos}, {Park}, {Parsons}, {Patel}, {Pen}, {Pesce}, {Pi{\'e}tu}, {Plambeck}, {PopStefanija}, {Porth}, {P{\"o}tzl}, {Prather}, {Preciado-L{\'o}pez}, {Psaltis}, {Pu}, {Ramakrishnan}, {Rao}, {Rawlings}, {Raymond}, {Rezzolla}, {Ricarte}, {Ripperda}, {Roelofs}, {Rogers}, {Romero-Ca{\~n}izales}, {Ros}, {Roshanineshat}, {Rottmann}, {Roy}, {Ruiz}, {Ruszczyk}, {Rygl}, {S{\'a}nchez}, {S{\'a}nchez-Arg{\"u}elles}, {S{\'a}nchez-Portal}, {Sasada}, {Satapathy}, {Savolainen}, {Schloerb}, {Schonfeld}, {Schuster}, {Shao}, {Shen}, {Small}, {Sohn}, {SooHoo}, {Sosapanta Salas}, {Souccar}, {Stanway}, {Sun}, {Tazaki}, {Tetarenko}, {Tiede}, {Tilanus}, {Titus}, {Torne}, {Toscano}, {Traianou}, {Trent}, {Trippe}, {Turk}, {van Bemmel}, {van Langevelde}, {van Rossum}, {Vos}, {Wagner}, {Ward-Thompson}, {Wardle}, {Washington}, {Weintroub}, {Wharton}, {Wielgus}, {Wiik}, {Witzel}, {Wondrak}, {Wong}, {Wu},
  {Yadlapalli}, {Yamaguchi}, {Yfantis}, {Yoon}, {Young}, {Young}, {Younsi}, {Yu}, {Yuan}, {Yuan}, {Zensus}, {Zhang}, {Zhao}, \& {Zhao}}]{2024ApJ...964L..25E}
{Event Horizon Telescope Collaboration}, {Akiyama}, K., {Alberdi}, A., {et~al.} 2024, ApJL, 964, L25

\bibitem[{{Event Horizon Telescope Collaboration} {et~al.}(2021){Event Horizon Telescope Collaboration}, {Akiyama}, {Algaba}, {Alberdi}, {Alef}, {Anantua}, {Asada}, {Azulay}, {Baczko}, {Ball}, {Balokovi{\'c}}, {Barrett}, {Benson}, {Bintley}, {Blackburn}, {Blundell}, {Boland}, {Bouman}, {Bower}, {Boyce}, {Bremer}, {Brinkerink}, {Brissenden}, {Britzen}, {Broderick}, {Broguiere}, {Bronzwaer}, {Byun}, {Carlstrom}, {Chael}, {Chan}, {Chatterjee}, {Chatterjee}, {Chen}, {Chen}, {Chesler}, {Cho}, {Christian}, {Conway}, {Cordes}, {Crawford}, {Crew}, {Cruz-Osorio}, {Cui}, {Davelaar}, {De Laurentis}, {Deane}, {Dempsey}, {Desvignes}, {Dexter}, {Doeleman}, {Eatough}, {Falcke}, {Farah}, {Fish}, {Fomalont}, {Ford}, {Fraga-Encinas}, {Friberg}, {Fromm}, {Fuentes}, {Galison}, {Gammie}, {Garc{\'\i}a}, {Gelles}, {Gentaz}, {Georgiev}, {Goddi}, {Gold}, {G{\'o}mez}, {G{\'o}mez-Ruiz}, {Gu}, {Gurwell}, {Hada}, {Haggard}, {Hecht}, {Hesper}, {Himwich}, {Ho}, {Ho}, {Honma}, {Huang}, {Huang}, {Hughes}, {Ikeda}, {Inoue}, {Issaoun},
  {James}, {Jannuzi}, {Janssen}, {Jeter}, {Jiang}, {Jimenez-Rosales}, {Johnson}, {Jorstad}, {Jung}, {Karami}, {Karuppusamy}, {Kawashima}, {Keating}, {Kettenis}, {Kim}, {Kim}, {Kim}, {Kim}, {Kino}, {Koay}, {Kofuji}, {Koch}, {Koyama}, {Kramer}, {Kramer}, {Krichbaum}, {Kuo}, {Lauer}, {Lee}, {Levis}, {Li}, {Li}, {Lindqvist}, {Lico}, {Lindahl}, {Liu}, {Liu}, {Liuzzo}, {Lo}, {Lobanov}, {Loinard}, {Lonsdale}, {Lu}, {MacDonald}, {Mao}, {Marchili}, {Markoff}, {Marrone}, {Marscher}, {Mart{\'\i}-Vidal}, {Matsushita}, {Matthews}, {Medeiros}, {Menten}, {Mizuno}, {Mizuno}, {Moran}, {Moriyama}, {Moscibrodzka}, {M{\"u}ller}, {Musoke}, {Mus Mej{\'\i}as}, {Michalik}, {Nadolski}, {Nagai}, {Nagar}, {Nakamura}, {Narayan}, {Narayanan}, {Natarajan}, {Nathanail}, {Neilsen}, {Neri}, {Ni}, {Noutsos}, {Nowak}, {Okino}, {Olivares}, {Ortiz-Le{\'o}n}, {Oyama}, {{\"O}zel}, {Palumbo}, {Park}, {Patel}, {Pen}, {Pesce}, {Pi{\'e}tu}, {Plambeck}, {PopStefanija}, {Porth}, {P{\"o}tzl}, {Prather}, {Preciado-L{\'o}pez}, {Psaltis}, {Pu},
  {Ramakrishnan}, {Rao}, {Rawlings}, {Raymond}, {Rezzolla}, {Ricarte}, {Ripperda}, {Roelofs}, {Rogers}, {Ros}, {Rose}, {Roshanineshat}, {Rottmann}, {Roy}, {Ruszczyk}, {Rygl}, {S{\'a}nchez}, {S{\'a}nchez-Arguelles}, {Sasada}, {Savolainen}, {Schloerb}, {Schuster}, {Shao}, {Shen}, {Small}, {Sohn}, {SooHoo}, {Sun}, {Tazaki}, {Tetarenko}, {Tiede}, {Tilanus}, {Titus}, {Toma}, {Torne}, {Trent}, {Traianou}, {Trippe}, {van Bemmel}, {van Langevelde}, {van Rossum}, {Wagner}, {Ward-Thompson}, {Wardle}, {Weintroub}, {Wex}, {Wharton}, {Wielgus}, {Wong}, {Wu}, {Yoon}, {Young}, {Young}, {Younsi}, {Yuan}, {Yuan}, {Zensus}, {Zhao}, \& {Zhao}}]{2021ApJ...910L..13E}
{Event Horizon Telescope Collaboration}, {Akiyama}, K., {Algaba}, J.~C., {et~al.} 2021, ApJL, 910, L13

\bibitem[{{Fragione}(2022)}]{2022ApJ...939...97F}
{Fragione}, G. 2022, ApJ, 939, 97

\bibitem[{{Fragione} {et~al.}(2022){Fragione}, {Kocsis}, {Rasio}, \& {Silk}}]{2022ApJ...927..231F}
{Fragione}, G., {Kocsis}, B., {Rasio}, F.~A., \& {Silk}, J. 2022, ApJ, 927, 231

\bibitem[{{Franchini} {et~al.}(2023){Franchini}, {Bonetti}, {Lupi}, {Miniutti}, {Bortolas}, {Giustini}, {Dotti}, {Sesana}, {Arcodia}, \& {Ryu}}]{2023A&A...675A.100F}
{Franchini}, A., {Bonetti}, M., {Lupi}, A., {et~al.} 2023, A\&A, 675, A100

\bibitem[{{Gilbaum} {et~al.}(2025){Gilbaum}, {Grishin}, {Stone}, \& {Mandel}}]{2025ApJ...982L..13G}
{Gilbaum}, S., {Grishin}, E., {Stone}, N.~C., \& {Mandel}, I. 2025, ApJL, 982, L13

\bibitem[{{GRAVITY Collaboration} {et~al.}(2018){GRAVITY Collaboration}, {Abuter}, {Amorim}, {Baub{\"o}ck}, {Berger}, {Bonnet}, {Brandner}, {Cl{\'e}net}, {Coud{\'e} Du Foresto}, {de Zeeuw}, {Deen}, {Dexter}, {Duvert}, {Eckart}, {Eisenhauer}, {F{\"o}rster Schreiber}, {Garcia}, {Gao}, {Gendron}, {Genzel}, {Gillessen}, {Guajardo}, {Habibi}, {Haubois}, {Henning}, {Hippler}, {Horrobin}, {Huber}, {Jim{\'e}nez-Rosales}, {Jocou}, {Kervella}, {Lacour}, {Lapeyr{\`e}re}, {Lazareff}, {Le Bouquin}, {L{\'e}na}, {Lippa}, {Ott}, {Panduro}, {Paumard}, {Perraut}, {Perrin}, {Pfuhl}, {Plewa}, {Rabien}, {Rodr{\'\i}guez-Coira}, {Rousset}, {Sternberg}, {Straub}, {Straubmeier}, {Sturm}, {Tacconi}, {Vincent}, {von Fellenberg}, {Waisberg}, {Widmann}, {Wieprecht}, {Wiezorrek}, {Woillez}, \& {Yazici}}]{2018A&A...618L..10G}
{GRAVITY Collaboration}, {Abuter}, R., {Amorim}, A., {et~al.} 2018, A\&A, 618, L10

\bibitem[{{Greene} {et~al.}(2020){Greene}, {Strader}, \& {Ho}}]{2020ARA&A..58..257G}
{Greene}, J.~E., {Strader}, J., \& {Ho}, L.~C. 2020, ARA\&A, 58, 257

\bibitem[{{Grishin} {et~al.}(2024){Grishin}, {Gilbaum}, \& {Stone}}]{2024MNRAS.530.2114G}
{Grishin}, E., {Gilbaum}, S., \& {Stone}, N.~C. 2024, MNRAS, 530, 2114

\bibitem[{{G{\"u}ltekin} \& {Miller}(2012)}]{2012ApJ...761...90G}
{G{\"u}ltekin}, K. \& {Miller}, J.~M. 2012, ApJ, 761, 90

\bibitem[{{Karas} {et~al.}(2017){Karas}, {Kop{\'a}{\v{c}}ek}, {Kunneriath}, {Zaja{\v{c}}ek}, {Araudo}, {Eckart}, \& {Kov{\'a}{\v{r}}}}]{2017CoSka..47..124K}
{Karas}, V., {Kop{\'a}{\v{c}}ek}, O., {Kunneriath}, D., {et~al.} 2017, Contributions of the Astronomical Observatory Skalnate Pleso, Vol.\ 47, pp.\ 124--132, arXiv.1705.09820

\bibitem[{{Kejriwal} {et~al.}(2024){Kejriwal}, {Witzany}, {Zaja{\v{c}}ek}, {Pasham}, \& {Chua}}]{2024MNRAS.532.2143K}
{Kejriwal}, S., {Witzany}, V., {Zaja{\v{c}}ek}, M., {Pasham}, D.~R., \& {Chua}, A. J.~K. 2024, MNRAS, 532, 2143

\bibitem[{{Kop{\'a}{\v{c}}ek} {et~al.}(2018){Kop{\'a}{\v{c}}ek}, {Tahamtan}, \& {Karas}}]{2018PhRvD..98h4055K}
{Kop{\'a}{\v{c}}ek}, O., {Tahamtan}, T., \& {Karas}, V. 2018, Physical Review D, 98, 084055

\bibitem[{{Linial} \& {Metzger}(2023)}]{2023ApJ...957...34L}
{Linial}, I. \& {Metzger}, B.~D. 2023, ApJ, 957, 34

\bibitem[{{Linial} \& {Metzger}(2024)}]{2024ApJ...963L...1L}
{Linial}, I. \& {Metzger}, B.~D. 2024, ApJL, 963, L1

\bibitem[{{Lipunov}(1992)}]{1992ans..book.....L}
{Lipunov}, V.~M. 1992, {Astrophysics of Neutron Stars} (in Astronomy and Astrophysics Library; Springer-Verlag Berlin, Heidelberg, New York)

\bibitem[{{Lu} \& {Quataert}(2023)}]{2023MNRAS.524.6247L}
{Lu}, W. \& {Quataert}, E. 2023, MNRAS, 524, 6247

\bibitem[{{Mathews}(1993)}]{1993ApJ...412L..17M}
{Mathews}, W.~G. 1993, Astrophysical Journal Letters, 412, L17

\bibitem[{{McKernan} {et~al.}(2014){McKernan}, {Ford}, {Kocsis}, {Lyra}, \& {Winter}}]{2014MNRAS.441..900M}
{McKernan}, B., {Ford}, K.~E.~S., {Kocsis}, B., {Lyra}, W., \& {Winter}, L.~M. 2014, MNRAS, 441, 900

\bibitem[{{McKernan} {et~al.}(2012){McKernan}, {Ford}, {Lyra}, \& {Perets}}]{2012MNRAS.425..460M}
{McKernan}, B., {Ford}, K.~E.~S., {Lyra}, W., \& {Perets}, H.~B. 2012, MNRAS, 425, 460

\bibitem[{{Miniutti} {et~al.}(2019){Miniutti}, {Saxton}, {Giustini}, {Alexander}, {Fender}, {Heywood}, {Monageng}, {Coriat}, {Tzioumis}, {Read}, {Knigge}, {Gandhi}, {Pretorius}, \& {Ag{\'\i}s-Gonz{\'a}lez}}]{2019Natur.573..381M}
{Miniutti}, G., {Saxton}, R.~D., {Giustini}, M., {et~al.} 2019, Nature, 573, 381

\bibitem[{{Olejak} {et~al.}(2025){Olejak}, {Stegmann}, {de Mink}, {Valli}, {Sari}, \& {Justham}}]{2025arXiv250321995O}
{Olejak}, A., {Stegmann}, J., {de Mink}, S.~E., {et~al.} 2025, arXiv e-prints, arXiv:2503.21995

\bibitem[{{Ostriker}(1999)}]{1999ApJ...513..252O}
{Ostriker}, E.~C. 1999, ApJ, 513, 252

\bibitem[{{Pasham} {et~al.}(2017){Pasham}, {Cenko}, {Sadowski}, {Guillochon}, {Stone}, {van Velzen}, \& {Cannizzo}}]{2017ApJ...837L..30P}
{Pasham}, D.~R., {Cenko}, S.~B., {Sadowski}, A., {et~al.} 2017, ApJL, 837, L30

\bibitem[{{Pasham} {et~al.}(2024{\natexlab{a}}){Pasham}, {Tombesi}, {Sukov{\'a}}, {Zaja{\v{c}}ek}, {Rakshit}, {Coughlin}, {Kosec}, {Karas}, {Masterson}, {Mummery}, {Holoien}, {Guolo}, {Hinkle}, {Ripperda}, {Witzany}, {Shappee}, {Kara}, {Horesh}, {van Velzen}, {Sfaradi}, {Kaplan}, {Burger}, {Murphy}, {Remillard}, {Steiner}, {Wevers}, {Arcodia}, {Buchner}, {Merloni}, {Malyali}, {Fabian}, {Fausnaugh}, {Daylan}, {Altamirano}, {Payne}, \& {Ferraraa}}]{2024SciA...10J8898P}
{Pasham}, D.~R., {Tombesi}, F., {Sukov{\'a}}, P., {et~al.} 2024{\natexlab{a}}, Science Advances, 10, eadj8898

\bibitem[{{Pasham} {et~al.}(2024{\natexlab{b}}){Pasham}, {Zaja{\v{c}}ek}, {Nixon}, {Coughlin}, {{\'S}niegowska}, {Janiuk}, {Czerny}, {Wevers}, {Guolo}, {Ajay}, \& {Loewenstein}}]{2024Natur.630..325P}
{Pasham}, D.~R., {Zaja{\v{c}}ek}, M., {Nixon}, C.~J., {et~al.} 2024{\natexlab{b}}, Nature, 630, 325

\bibitem[{{Payne} {et~al.}(2022){Payne}, {Shappee}, {Hinkle}, {Holoien}, {Auchettl}, {Kochanek}, {Stanek}, {Thompson}, {Tucker}, {Armstrong}, {Boyd}, {Brimacombe}, {Cornect}, {Huber}, {Jha}, \& {Lin}}]{2022ApJ...926..142P}
{Payne}, A.~V., {Shappee}, B.~J., {Hinkle}, J.~T., {et~al.} 2022, ApJ, 926, 142

\bibitem[{{Peters}(1964)}]{1964PhRv..136.1224P}
{Peters}, P.~C. 1964, Physical Review, 136, 1224

\bibitem[{{Ripperda} {et~al.}(2022){Ripperda}, {Liska}, {Chatterjee}, {Musoke}, {Philippov}, {Markoff}, {Tchekhovskoy}, \& {Younsi}}]{2022ApJ...924L..32R}
{Ripperda}, B., {Liska}, M., {Chatterjee}, K., {et~al.} 2022, ApJL, 924, L32

\bibitem[{{Rose} {et~al.}(2022){Rose}, {Naoz}, {Sari}, \& {Linial}}]{2022ApJ...929L..22R}
{Rose}, S.~C., {Naoz}, S., {Sari}, R., \& {Linial}, I. 2022, ApJL, 929, L22

\bibitem[{{Sillanpaa} {et~al.}(1988){Sillanpaa}, {Haarala}, {Valtonen}, {Sundelius}, \& {Byrd}}]{1988ApJ...325..628S}
{Sillanpaa}, A., {Haarala}, S., {Valtonen}, M.~J., {Sundelius}, B., \& {Byrd}, G.~G. 1988, ApJ, 325, 628

\bibitem[{{Sniegowska} {et~al.}(2020){Sniegowska}, {Czerny}, {Bon}, \& {Bon}}]{2020A&A...641A.167S}
{Sniegowska}, M., {Czerny}, B., {Bon}, E., \& {Bon}, N. 2020, A\&A, 641, A167

\bibitem[{{{\'S}niegowska} {et~al.}(2023){{\'S}niegowska}, {Grz{\c{e}}dzielski}, {Czerny}, \& {Janiuk}}]{2023A&A...672A..19S}
{{\'S}niegowska}, M., {Grz{\c{e}}dzielski}, M., {Czerny}, B., \& {Janiuk}, A. 2023, A\&A, 672, A19

\bibitem[{{Stone} \& {Metzger}(2016)}]{2016MNRAS.455..859S}
{Stone}, N.~C. \& {Metzger}, B.~D. 2016, MNRAS, 455, 859

\bibitem[{{Stone} \& {van Velzen}(2016)}]{2016ApJ...825L..14S}
{Stone}, N.~C. \& {van Velzen}, S. 2016, ApJL, 825, L14

\bibitem[{{Sukov{\'a}} {et~al.}(2024){Sukov{\'a}}, {Tombesi}, {Pasham}, {Zaja{\v{c}}ek}, {Wevers}, {Ryu}, {Linial}, \& {Franchini}}]{2024arXiv241104592S}
{Sukov{\'a}}, P., {Tombesi}, F., {Pasham}, D.~R., {et~al.} 2024, arXiv e-prints, arXiv:2411.04592

\bibitem[{{Sukov{\'a}} {et~al.}(2023){Sukov{\'a}}, {Zaja{\v{c}}ek}, \& {Karas}}]{2023arXiv231204149S}
{Sukov{\'a}}, P., {Zaja{\v{c}}ek}, M., \& {Karas}, V. 2023, in Proceedings of RAGtime 23--25 Workshop (Silesian University, Opava), 109--117, arXiv:2312.04149

\bibitem[{{Sukov{\'a}} {et~al.}(2021){Sukov{\'a}}, {Zaja{\v{c}}ek}, {Witzany}, \& {Karas}}]{2021ApJ...917...43S}
{Sukov{\'a}}, P., {Zaja{\v{c}}ek}, M., {Witzany}, V., \& {Karas}, V. 2021, ApJ, 917, 43

\bibitem[{{Syer} {et~al.}(1991){Syer}, {Clarke}, \& {Rees}}]{1991MNRAS.250..505S}
{Syer}, D., {Clarke}, C.~J., \& {Rees}, M.~J. 1991, MNRAS, 250, 505

\bibitem[{{Toropina} {et~al.}(2012){Toropina}, {Romanova}, \& {Lovelace}}]{2012MNRAS.420..810T}
{Toropina}, O.~D., {Romanova}, M.~M., \& {Lovelace}, R.~V.~E. 2012, Mon. Not. Roy. Astronon. Soc., 420, 810

\bibitem[{{{\v{S}}tolc} {et~al.}(2023){{\v{S}}tolc}, {Zaja{\v{c}}ek}, {Czerny}, \& {Karas}}]{2023MNRAS.522.2869S}
{{\v{S}}tolc}, M., {Zaja{\v{c}}ek}, M., {Czerny}, B., \& {Karas}, V. 2023, MNRAS, 522, 2869

\bibitem[{{{\v{S}}ubr} {et~al.}(2004){{\v{S}}ubr}, {Karas}, \& {Hur{\'e}}}]{2004MNRAS.354.1177S}
{{\v{S}}ubr}, L., {Karas}, V., \& {Hur{\'e}}, J.~M. 2004, Mon. Not. Roy. Astronon. Soc., 354, 1177

\bibitem[{{Yuan} \& {Narayan}(2014)}]{2014ARA&A..52..529Y}
{Yuan}, F. \& {Narayan}, R. 2014, ARA\&A, 52, 529

\bibitem[{{Zaja{\v{c}}ek} {et~al.}(2024){Zaja{\v{c}}ek}, {Czerny}, {Jaiswal}, {{\v{S}}tolc}, {Karas}, {Pandey}, {Pasham}, {{\'S}niegowska}, {Witzany}, {Sukov{\'a}}, {M{\"u}nz}, {Werner}, {{\v{R}}{\'\i}pa}, {Merc}, {Labaj}, {Kurf{\"u}rst}, \& {Krti{\v{c}}ka}}]{2024SSRv..220...29Z}
{Zaja{\v{c}}ek}, M., {Czerny}, B., {Jaiswal}, V.~K., {et~al.} 2024, Space Science Reviews, 220, 29

\bibitem[{{Zaja{\v{c}}ek} {et~al.}(2015){Zaja{\v{c}}ek}, {Karas}, \& {Kunneriath}}]{2015AcPol..55..203Z}
{Zaja{\v{c}}ek}, M., {Karas}, V., \& {Kunneriath}, D. 2015, Acta Polytechnica, 55, 203

\bibitem[{{Zaja{\v{c}}ek} {et~al.}(2025){Zaja{\v{c}}ek}, {Werner}, {Best}, {Esme L'Heureux}, {{\v{R}}{\'\i}pa}, {Pikhartov{\'a}}, {Mondek}, {M{\"u}nz}, {{\v{S}}tofanov{\'a}}, {Kurf{\"u}rst}, {Labaj}, {Garland}, {Tohuvavohu}, {Karas}, \& {Sukov{\'a}}}]{2025arXiv250119365Z}
{Zaja{\v{c}}ek}, M., {Werner}, N., {Best}, H., {et~al.} 2025, arXiv e-prints, arXiv:2501.19365

\bibitem[{{Zurek} {et~al.}(1992){Zurek}, {Siemiginowska}, \& {Colgate}}]{1992AIPC..254..564Z}
{Zurek}, W.~H., {Siemiginowska}, A., \& {Colgate}, S.~A. 1992, in American Institute of Physics Conference Series, Vol. 254, Testing the AGN paradigm, ed. S.~S. {Holt}, S.~G. {Neff}, \& C.~M. {Urry} (AIP), pp.\ 564--567

\end{thebibliography}

\end{document}